\documentclass[preprint]{aastex63}

\usepackage{savesym}
\savesymbol{tablenum}
\usepackage{siunitx}
\restoresymbol{SIX}{tablenum}

\DeclareSIUnit\parsec{pc}
\DeclareSIUnit\lightyear{ly}

\let\OldAng\ang%
\renewcommand*{\ang}[2][]{%
    \OldAng[scientific-notation=false,
    separate-uncertainty=true,#1]{#2}%
}
\DeclareSIUnit\year{yr}
\DeclareSIUnit\erg{erg}
\DeclareSIUnit\msun{M_{\astrosun}}
\DeclareSIUnit{\GeV}{\giga\electronvolt}
\DeclareSIUnit{\TeV}{\tera\electronvolt}
\DeclareSIUnit{\PeV}{\peta\electronvolt}
\DeclareSIUnit{\MeV}{\mega\electronvolt}
\DeclareSIUnit{\eV}{\electronvolt}
\DeclareSIUnit{\smm}{\square\metre\second}
\DeclareSIUnit{\smmr}{\metre^{-2}\second^{-1}}
\DeclareSIUnit{\pe}{p.e.}

\usepackage[utf8]{inputenc}
\usepackage{tabularx}
\usepackage{booktabs}

\usepackage[capitalize,noabbrev]{cleveref}

\newcommand{\theSource}{HAWC~J2227+610}
\newcommand{\theSnr}{G106.3+2.7}
\newcommand{\e}[1]{\ensuremath{\cdot 10^{#1}}}

\shorttitle{HAWC J2227+610}

\begin{document}
\title{HAWC J2227+610 and its association with G106.3+2.7, a new potential Galactic PeVatron}

\correspondingauthor{Henrike Fleischhack}
\email{hfleisch@mtu.edu}



\author[0000-0003-0197-5646]{A.~Albert}
\affiliation{Physics Division, Los Alamos National Laboratory, Los Alamos, NM, USA }
\author[0000-0001-8749-1647]{R.~Alfaro}
\affiliation{Instituto de F\'{i}sica, Universidad Nacional Autónoma de México, Ciudad de Mexico, Mexico }
\author{C.~Alvarez}
\affiliation{Universidad Autónoma de Chiapas, Tuxtla Gutiérrez, Chiapas, México}
\author{J.R.~Angeles Camacho}
\affiliation{Instituto de F\'{i}sica, Universidad Nacional Autónoma de México, Ciudad de Mexico, Mexico }
\author{J.C.~Arteaga-Velázquez}
\affiliation{Universidad Michoacana de San Nicolás de Hidalgo, Morelia, Mexico }
\author{K.P.~Arunbabu}
\affiliation{Instituto de Geof\'{i}sica, Universidad Nacional Autónoma de México, Ciudad de Mexico, Mexico }
\author{D.~Avila Rojas}
\affiliation{Instituto de F\'{i}sica, Universidad Nacional Autónoma de México, Ciudad de Mexico, Mexico }
\author[0000-0002-2084-5049]{H.A.~Ayala Solares}
\affiliation{Department of Physics, Pennsylvania State University, University Park, PA, USA }
\author[0000-0003-0477-1614]{V.~Baghmanyan}
\affiliation{Institute of Nuclear Physics Polish Academy of Sciences, PL-31342 IFJ-PAN, Krakow, Poland }
\author[0000-0003-3207-105X]{E.~Belmont-Moreno}
\affiliation{Instituto de F\'{i}sica, Universidad Nacional Autónoma de México, Ciudad de Mexico, Mexico }
% ...

\collaboration{100}{(HAWC Collaboration)}


\begin{abstract}
We present the detection of VHE gamma-ray emission above \SI[scientific-notation=engineering]{100}{\TeV} from HAWC J2227+610 with the HAWC observatory. Combining our observations with previously published results by VERITAS, we interpret the gamma-ray emission from HAWC J2227+610 as emission from protons with a lower limit in their cutoff energy of 800\,TeV. The most likely source of the protons is the associated supernova remnant G106.3+2.7, making it a good candidate for a Galactic PeVatron. However, a purely leptonic origin of the observed emission cannot be excluded at this time.
\end{abstract}



\section{Introduction} \label{intro}

Although their existence has been known for more than hundred years \citep{Hess:1912srp}, the origins of cosmic rays are still not fully understood. These charged particles, mostly protons and fully-ionized nuclei, of extra-terrestrial origin have been detected over many orders of magnitude in energy \citep{Tanabashi:2018oca}. According to our current understanding, cosmic rays with energies up to a few PeV (the ``knee'', a softening in the measured energy spectrum) are thought to originate from sources (accelerators) within our own Galaxy. But what are those sources? Supernova remnants have been proposed as potential sources of Galactic cosmic rays, mainly for two reasons (see e.g. \citep{BELL201356} and references therein). First, the diffusive shock acceleration mechanism provides an efficient way to transfer energy from the exploding superova shell into cosmic rays and accelerate them to relativistic energies. And second, with a rate of a few supernovae per century, SNRs have a sufficient energy budget to accelerate the bulk of the Galaxy's cosmic rays. 

As cosmic rays are deflected by Galactic magnetic fields, their arrival directions at Earth generally do not point back to their souces. However, relativistic protons interacting with gas and dust near their sources can produce GeV--TeV gamma-ray emission. And indeed, at least two SNRs have gamma-ray spectra with a characteristic ``pion bump'' feature, indicating that they accelerate protons to relativistic energies \citep{Ackermann:2013wqa}. But so far, no SNR has shown to emit gamma rays to hundreds of TeV as would be indicative of a PeVatron (a source capable of accelerating protons to at least PeV energies or higher).

Nonetheless, the search for PeVatrons has not been fruitless. H.E.S.S. has detected evidence for the existence of a PeVatron near the Galactic center \citep{Abramowski:2016mir}, although the source of the high-energy protons there is now yet known. HAWC recently performed a blind search for gamma-ray emission above 100\,TeV \citep{PhysRevLett.124.021102}. All three $>$100\, TeV sources identified in that study show the characteristic spectral curvature indicative of gamma-ray emission from relativistic electrons rather than protons, and  they all have energetic pulsars nearby, which are potential soures of relativistic electrons. Still, more in-depth studies regarding the nature of these sources are in progress.

The supernova remnant (SNR) \theSnr{} is a comet-shaped radio source, with a brighter `head' and an extended `tail' region \citep{1990A&AS...82..113J, Pineault_2000}. The pulsar PSR J2229+6114, seen in radio, X-rays, and gamma rays \citep{Hartman_1999, Halpern_2001, Abdo_2009}, and its pulsar wind nebula (PWN) G106.65+2.96, the Boomerang Nebula \citep{Halpern_2001, Kothes_2006}, appear to be contained inside the remnant. 




Measurements of the gas structure in the region put PSR~J2229+6114 and \theSnr{} at a distance of about \SI[scientific-notation=engineering]{800}{\parsec} from Earth \citep{Kothes_2001}; closer than originally thought. These measurements suggest that PSR J2229+6114 and \theSnr{} were created from the same progenitor event, with the asymmetric structure resulting from inhomogeneities in the medium surrounding the progenitor star. 



VHE (very-high-energy, $E>\SI[scientific-notation=engineering]{100}{\GeV}$) emission from this region has been reported by the Milagro collaboration at 20\,TeV \citep{Abdo_2007} and 35\,TeV \citep{Abdo_2009_MGRO, Abdo_2009_erratum, MGRO_Atel}, and by the VERITAS  collaboration in the energy range from 900\,GeV to 16\,TeV \citep{2009ApJ...703L...6A}. The Milagro source (MGRO J2228+61) is spatially extended and consistent with both the pulsar position and the radio `tail' within uncertainties. The VERITAS source is spatially asymmetric. The centroid of the emission is offset from the pulsar, consistent with a molecular cloud in the radio `tail', and consistent with the Milagro source within uncertainties. The emission detected by VERITAS follows a hard power-law energy spectrum ($\propto E^{-2.3}$). Milagro reported measurements of the differential flux at \SI[scientific-notation=engineering]{20}{\TeV} and  \SI[scientific-notation=engineering]{35}{\TeV}, which are consistent with the extrapolation of the VERITAS spectrum to higher energies within uncertainties. 

While PSR J2229+6114 is a well-established GeV gamma-ray source, there is no clear GeV counterpart of \theSnr{} \citep{Acero_2016, Ackermann_2017, Fermi-LAT:2019yla}. Recently, an independent group of authors analysed 10 years of Fermi-LAT from 3\,GeV to 500\,GeV and reported a weak, but significantly detected, GeV gamma-ray source with a hard energy spectrum. This GeV source is best described by a disk morphology with a radius of \ang{0.25}, overlapping with the VHE emission region \citep{2019ApJ...885..162X}. 

 
\cite{2019ApJ...885..162X} also show models of the radio and (V)HE gamma ray emission from \theSnr{}. Their study prefers a lepto-hadronic model --- with the gamma-ray emission dominated by hadronic processes and the proton energy spectrum extending to at least \SI[scientific-notation=engineering]{400}{\TeV} --- over a purely leptonic model.

The previous detections of gamma-ray emission up to tens of TeV make \theSnr{} a potential \emph{Galactic pevatron}: a source that would be able to accelerate cosmic rays up to PeV energies. It would only be the second such source in our Galaxy.


In this paper, we report on the HAWC detection of multi-TeV emission coincident with \theSnr{}. Section \ref{results} describes the main results, including constraints derived on the underlying particle population, for two models, assuming either a hadronic or a leptonic origin of the VHE emission. The potential sources of the particles producing the gamma-ray emission are discussed in Section \ref{discussion}. Future prospects are summarized in Section \ref{conclusions}. Details on the data analysis can be found in Section \ref{analysis} in the supplemental materials.
\section{Results}
\label{results}

\begin{figure}[tb]
\includegraphics[width=0.5\textwidth]{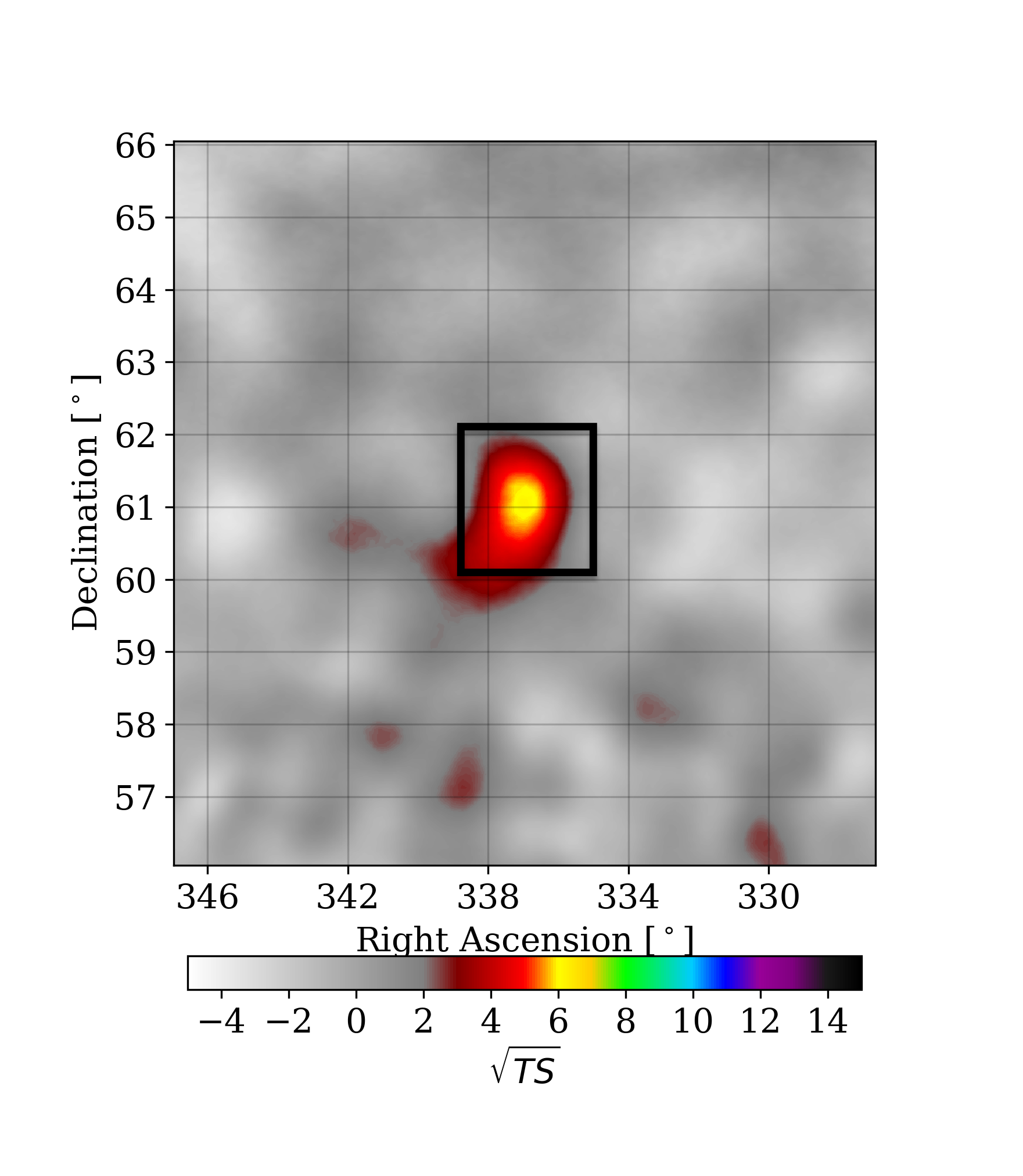}\includegraphics[width=0.5\textwidth]{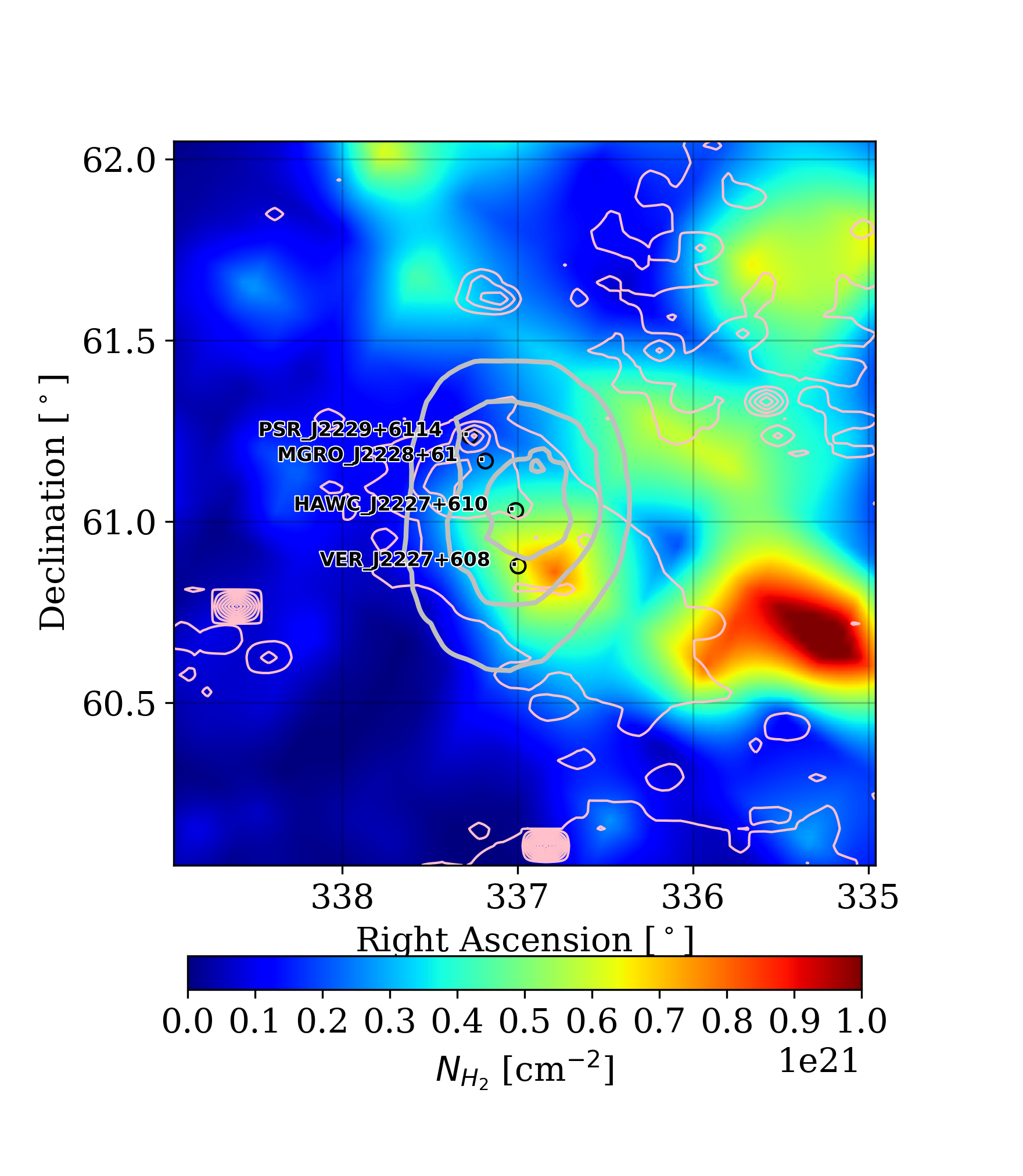}
\caption{Left: HAWC significance map of the region, large scale view. There are no other significant gamma-ray sources nearby that could affect the measurement. The black frame marks the size of the region shown on the right. Right: Molecular hydrogen column density around \theSource{}. See Section \ref{molecular_gas} in the supplemental materials for more details. The pulsar  position as well as the centroids of the VERITAS and Milagro sources have been marked. The grey contours show the 1\,$\sigma$, 2\,$\sigma$, 3\,$\sigma$ confidence regions for the HAWC source position. The pink contours show the \SI{1.4}{\giga\hertz} continuum brightness temperature from the Canadian Galactic Plane Survey \citep{CGPS} in 50 logarithmically spaced steps from \SI{1}{\kelvin} to \SI[exponent-to-prefix=false, scientific-notation=false]{100}{\kelvin}. Both maps have been smoothed and interpolated for display.}  
\label{fig:sigmap}
\end{figure}

\subsection{Source search}
The blind point source search described in Section~\ref{analysis} found a local maximum at RA=\ang{336.96}, Dec=\ang{61.05}, with a statistical uncertainty of \ang{0.16} and a systematic uncertainty of \ang{0.1}. The significance map for the search is shown in \cref{fig:sigmap}. The excess is well isolated and is inconsistent with background fluctuations at the 6.2\,$\sigma$ level (pre-trials), or about 4.3\,$\sigma$ post-trials considering HAWC's entire field of view. It is designated as a new source, \theSource{}.
 
\subsection{Morphology}
The best-fit position of \theSource{} is consistent with the VHE detections by VERITAS and Milagro within uncertainties (see \cref{fig:sigmap}). It is also consistent with the pulsar position.

In addition to the point source hypothesis, \theSource{} is fit with a symmetric Gaussian morphology, with the width left free. The extended model is not prefered over the point source hypothesis. (The faint `tail' towards the southeast in \cref{fig:sigmap} is not statistically significant.) At 90\% confidence level (CL), the Gaussian extent is constrained to be less than $\left(0.232^{+0.024}_{-0.004}\left(\mathrm{syst.}\right)\right)$\si{\degree}. The HAWC morphology is consistent with the asymmetric Gaussian morphology measured by VERITAS.

As the morphology of the VERITAS source and the HAWC source are consistent with each other, we assume that the VHE emission seen by the two instruments is due to the same particle population/underlying emission mechanism.

\subsection{Gamma-ray energy spectrum}
\begin{table}[tp]
\caption{Spectral fit parameters for each dataset as well as for the joint fit. VERITAS results are taken from \cite{2009ApJ...703L...6A} (not adjusted emission outside the VERITAS integration region) and shown here for convenience only. For the power-law flux normalization and spectral index, both statistical and systematic uncertainties are given. For the lower limit on the cutoff energy, only systematic uncertainties are given.} 
\label{tab:fit}
\begin{center}
\begin{tabular*}{\textwidth}{@{\extracolsep{\fill} } lcccc}
\toprule
 &  VERITAS & HAWC & Joint fit \\ \midrule
 $E_0$  [TeV] & 3 & 80 & 20 \\
 $K$ [TeV$^{-1}$cm$^{-2}$s$^{-1}$] & 
 $\left( 1.15\pm 0.27 \pm 0.35\right) \e{-13}$ &
 $\left( 1.02 \pm 0.17 ^{+0.17}_{-0.22}\right) \e{-16}$ &
 $\left( 2.46 \pm 0.35 ^{+0.33}_{-0.47}\right) \e{-15}$ \\
 $\gamma$  & $-2.29\pm 0.33\pm 0.30$ &
  $-2.25\pm 0.23 ^{+0.03}_{-0.19}$ &
  $-2.29\pm 0.08 \pm 0.09 $ \\
 $E_{min}$  [TeV] &  0.9 & 40 & 0.9 \\
 $E_{max}$  [TeV] & 17 & 110 & 180 \\ \midrule
 $E_C$ (90\% CL) [TeV] & & $>35.7^{+0.1}_{-16.9}$ & $>120^{+81}_{-76}$ \\ \bottomrule
\end{tabular*}
\end{center}
\end{table}



\begin{figure}[bt]
\includegraphics[width=\textwidth]{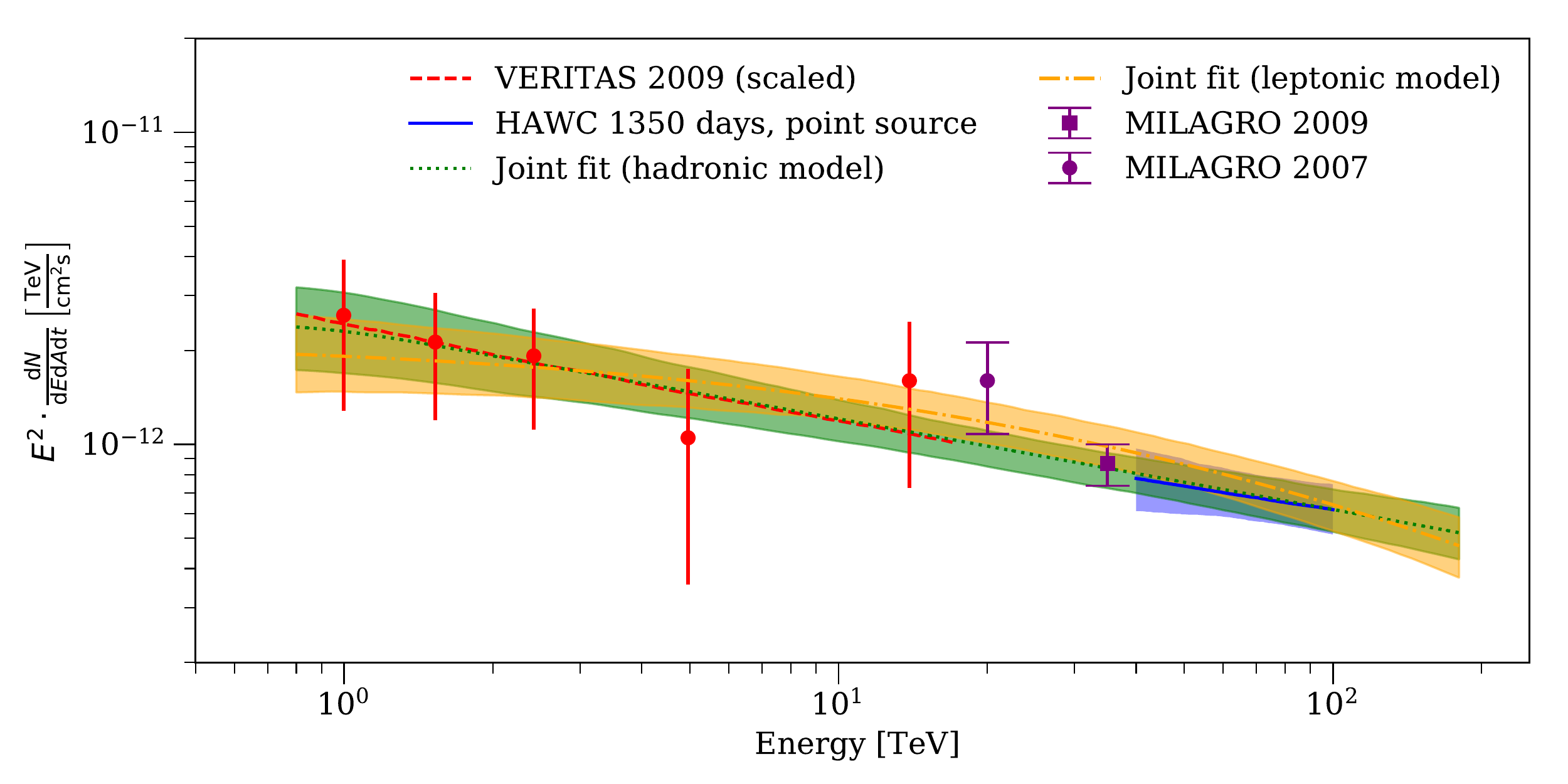}
\caption{VHE gamma-ray energy spectrum of \theSource{}, statistical uncertainties only. VERITAS data points and fit results from \cite{2009ApJ...703L...6A} have been scaled to account for the difference between the size of the emission region and the region over which the spectrum was extracted. Two joint fits to HAWC and VERITAS data are shown: One assuming a pion decay spectrum from a proton population following a power-law energy spectrum (hadronic model), and one assuming the gamma-ray emission is due to inverse Compton emission from electron (leptonic model). Milagro data points from \cite{Abdo_2007, Abdo_2009_MGRO} are shown for reference only.}
\label{fig:spectra}
\end{figure}

Considering only HAWC data, the energy spectrum of \theSource{} is well fit by a power law. Including spectral curvature or a cutoff does not significantly improve the likelihood. The fit parameters and lower limits on the cutoff energy are given in \cref{tab:fit}. The best-fit spectral index agrees well with the VERITAS measurement. After scaling the VERITAS data points to account for the emission outside the VERITAS integration region (c.f. section \ref{sec:data_analysis} in the appendix), the energy spectrum measured by HAWC lines up very well with the extrapolation of the VERITAS measurement to higher energies, as well as with the Milagro flux points (see \cref{fig:spectra}), without any indication of a break or a cutoff. 

The joint VERITAS-HAWC spectrum is well fit by a power law from \SI[scientific-notation=engineering]{0.9}{\TeV} to \SI[scientific-notation=engineering]{180}{\TeV}. At 90\% CL, a lower limit on the cutoff energy is placed at $120^{+81}_{-60}\left(\mathrm{syst.}\right)$\,\si{\TeV} (assuming an exponential cutoff).

\subsection{Primary proton energy spectrum}


The VHE gamma-ray emission from \theSource{} is modeled as a pion decay spectrum, assuming that the underlying proton spectrum follows a power law. The best-fit spectral index of the relativistic proton population is given by $\gamma_p=-2.35 \pm 0.07 \left(\mathrm{stat.}\right) ^{+0.09}_{-0.10}\left(\mathrm{syst.}\right)$. The best-fit normalization (total proton energy above \SI{2}{\TeV} times gas density) is given by \\$n \, W_p = \left( 4.4 \pm 0.9 \left(\mathrm{stat.}\right) ^{+1.3}_{-1.2}\left(\mathrm{syst.}\right)\right)\e{48}\,\si{\erg\per\centi\metre\cubed}\cdot\left(d/\left(800\,\si{\parsec}\right)\right)^2$, where $d$ is the distance to the source. 

An exponential cutoff in the proton spectrum does not significantly improve the likelihood. The $90\%$ lower limit on the proton cutoff energy is given by $E_{C,p} > 800 ^{+990}_{-640}\left(\mathrm{syst.}\right)\,\si{\TeV}$.

\subsection{Primary electron energy spectrum}
\label{leptonic}


The model was optimized using the same maximum-likelihood procedure as for the hadronic case. The best-fit values of the parameters are $W_e = \left( 4.2 \pm 1.2 \left( \textrm{stat.}\right)^{+1.6}_{-1.5} \left( \textrm{syst.}\right)  \right)\cdot 10^{45}\,\si{\erg}\cdot\left(d/\left(800\,\si{\parsec}\right)\right)^2$ and $\gamma_e=-2.75 \pm 0.11\left( \textrm{stat.}\right) ^{+0.15}_{-0.14}\left(\textrm{syst.}\right)$, where $d$ is the distance to the source. The resulting predicted gamma-ray energy spectrum can be seen in \cref{fig:spectra}. No evidence for curvature or a cutoff in the electron spectrum was found. The 90\% lower limit on the electron cutoff energy is given by $E_{C,e} > 270 ^{+140}_{-170}\left(\mathrm{syst.}\right)$\,TeV.


The GHz radio emission seen from \theSnr{} (see e.g. \cite{Pineault_2000}) could be explained by synchrotron emission from electrons with energies in the GeV range. Without precise knowledge of the B-field in the region, the radio and TeV data alone are not sufficient to constrain the electron spectrum in the GeV to TeV range. Better constraints on the synrchrotron peak, e.g. from X-ray measurements, would be needed. Due to the relatively large size of the emission region, multi-wavelenght coverage is rather sparse. For this study, no attempt was made to model the emission at radio wavelengths or any other energy range other than TeV gamma rays.

\clearpage


\section{Origin of the VHE gamma-ray emission}
\label{discussion}





Using VHE gamma-ray data alone, we cannot distinguish whether the gamma-ray emission is due to leptonic or hadronic processes. Both the best-fit leptonic and the best-fit hadronic models for \theSnr{} have similar values for the optimized likelihood. We use the Bayesian Information Criterion \citep{schwarz1978} to compare the two models. We find $\Delta BIC = BIC_{\mathrm{leptonic}} - BIC_{\mathrm{hadronic}} = 0.34$ (in favor of the hadronic model), which does not indicate that one of the models is significantly preferred over the other. 

Observations at other wavelengths will be needed to ascertain the true origin of the gamma-ray emission. If the observed radiation is of hadronic origin, one might be able to observe a pion-bump signature at hundreds of MeV \citep{Ackermann:2013wqa}. On the other hand, if the emission is of leptonic origin, the same electrons are expected to also emit synchrotron emission up to X-ray energies (depending on the ambient magnetic field).

\subsection{Hadronic Origin}
Assuming a hadronic origin of the observed gamma-ray emission, the underlying population of relativistic protons is described by a hard power-law energy spectrum, with spectral index $-2.35$ above \SI{2}{\TeV} and no cutoff below at least \SI[scientific-notation=engineering]{800}{\TeV}. What could be the source of these protons?

Assuming a gas density of $n\approx \SI[scientific-notation=false, exponent-to-prefix=false]{50}{\per\cubic\centi\metre}$ (see Appendix \ref{molecular_gas}), the best-fit normalization of the proton spectrum corresponds to a total energy in protons above \SI[scientific-notation=engineering]{2}{\TeV} of $W_p=\left(9 \pm 2\left(\mathrm{stat.}\right)\pm 3\left(\mathrm{syst.}\right)\right)\e{46}\,\si{\erg}$. Assuming that the spectrum extends down to \SI[scientific-notation=engineering]{1}{\GeV} without a break, this corresponds to a total energy in protons above \SI[scientific-notation=engineering]{1}{\GeV} of $W_{p,>\SI[scientific-notation=engineering]{1}{\GeV}} \approx \SI[scientific-notation=true, exponent-to-prefix=false]{1e48}{\erg}$, within a factor of two (large uncertainties as no GeV data were considered for this measurement). 

\cite{2019ApJ...885..162X} found a total energy in protons above \SI[scientific-notation=engineering]{1}{\GeV} of $\SI[exponent-to-prefix=false]{6e48}{\erg}$ for $n=\SI{1}{\per\centi\metre\cubed}$, which would correspond to $W_{p,>\SI[scientific-notation=engineering]{1}{\GeV}} \approx \SI[exponent-to-prefix=false]{1.2e47}{\erg}$ assuming $n\approx \SI[scientific-notation=false, exponent-to-prefix=false]{50}{\per\cubic\centi\metre}$; almost an order of magnitude lower than the HAWC-VERITAS result. This discrepancy can be explained by the fact that they included a new GeV gamma-ray source in the region which they ascribe to the same proton population. Including the GeV emission, their model prefers a harder index of $-2$ for the proton spectrum and thus contains fewer protons at lower energies.

The most likely source of these protons is diffusive shock acceleration in the SNR \theSnr{}, which has sufficient energy budget assuming even a relatively small acceleration efficiency of about 1\% (\cite{Kothes_2001} derived a lower limit to the total kinetic energy of the SNR shock of about \SI[exponent-to-prefix=false]{7e49}{\erg}). 

However, SNRs are generally only expected to act as PeVatrons for the first few hundred years of their evolution \citep{AHARONIAN201371}, while \theSnr{} could be as old as ten thousand years. We offer two possible scenarios that could explain the observed VHE emission.

\subsubsection{Old SNR Scenario}
Assuming that the characteristic age of PSR J2229+6114 (10.4\,\si{\kilo\year}) is close to the real age of the system, \theSnr{} is not expected to be an active proton accelerator up to hundreds of TeV at this stage of it evolution. The observed VHE emission would be due to protons previously accelerated to hundreds of TeV by the SNR, which are now diffusing freely in the region and interacting with a molecular cloud outside the accelerating region.

The energy-dependence of the diffusion coefficient $D(E_p)$ of protons in the interstellar medium (ISM) can be approximated by a power-law: $D(E_p) = D_0 \, \left( E_p/ \si{GeV} \right)^\delta$, with $D_0 \approx 3\e{24}\,\si{\centi\metre\squared\per\second}$ and $\delta\approx 1$ for Kolmogorov diffusion \citep{Aharonian2012}.

The characteristic length scale that a diffusing particle of energy $E$ travels in time $t$ is given by $l=2\, \sqrt{t\, D(E)}$ \citep{PhysRevD.52.3265}. Assuming that the source was mainly active for the first few hundred years of its evolution, then $t=10\,\si{\kilo\year}$, which yields $l\approx 600\,\si{\parsec}$ for an 800\,TeV proton (and 200\,\si{\parsec} for a 2\,TeV proton). This distance is much larger than the few pc size of the emission region, indicating that most of the protons accelerated by the SNR should already have diffused much further out into the Galaxy. However, it is expected that diffusion is suppressed by a factor of 100 or more near SNRs \citep{Fujita_2009}, corresponding to a diffusion length of $60\,\si{\parsec}$ or less for an 800\,TeV proton, and enough protons could remain to produce the observed emission. 

In the case of an ``old'' SNR, the total energy in relativistic particles would have to be higher than derived above, as only a fraction of the accelerated protons would have diffused into the emission region. The population of freshly accelerated protons must have had a harder spectral index than the one measured here, since higher-energy protons would have had a larger chance of diffusing away from the emission zone.

\subsubsection{Young SNR scenario}
The characteristic age of a pulsar is only a valid  approximation for its true age if its current spin period is much slower than its initial period, and its braking index is close to 3. This is not necessarily the case for PSR J2229+6114 with its current period of about \SI[scientific-notation=engineering]{51.6}{\milli\second} \citep{Halpern_2001}, as initial periods of even \emph{more} than \SI[scientific-notation=engineering]{50}{\milli\second} are not uncommon \citep{10.1093/mnras/stt047}. Assuming its initial period was close to the current value, PSR J2229+6114 could be much younger than its characteristic age. On the other hand, PSR J2229+6114 is also thought to be the source of the Boomerang PWN G106.6+2.9, which has been estimated to be at least \SI[scientific-notation=false]{3900}{\year} old \citep{Kothes_2006}.

There is, however, still a possibility that the SNR \theSnr{} is not connected to PSR J2229+6114 and merely happens to lie on the same line of sight as the Boomerang PWN. In that case, the age of \theSnr{} is not well constrained, although is is likely to be at least a few hundred years old as there are no historic records of the supernova explosion.

If the SNR is only hundreds of years old, it could still be an active particle accelerator, and the gamma-ray emission would be due to freshly accelerated protons interacting with molecular gas inside or very nearby the acceleration region. 







\subsubsection{\theSource{} as a potential neutrino emitter}
Assuming that the VHE gamma-ray emission from \theSource{} is indeed dominated by the decays of neutral pions from $pp$ collisions, it should be a source of VHE neutrinos from the decay of charged pions, which are also produced in $pp$ interactions. These neutrinos could then be detected by suitable detectors such as IceCube \citep{2013Sci...342E...1I} or ANTARES \citep{AGERON201111}. However, neither experiment has detected significant \emph{steady} neutrino point sources yet \citep{Aartsen:2018ywr, PhysRevD.96.082001, Illuminati:2019oag}.

According to \cite{Ahlers:2013xia}, the flux of muon neutrinos (including anti-neutrinos) should be proportional to the gamma-ray flux and is approximated by 
\begin{equation}
\frac{dN_{\nu_\mu}}{dE_{\nu_\mu}} = \frac{E_\gamma}{E_{\nu_\mu}} \, \frac{dN_\gamma}{dE_\gamma}, 
\end{equation}
where the neutrino energy $E_\nu$ and the gamma-ray energy $E_\gamma$ are related as $E_\gamma \approx 2\, E_\nu$. Here, we neglected any gamma-ray absorption effects, and assumed equal flavor ratios at Earth due to flavor mixing. Only muon (anti-)neutrinos are considered here as IceCube has published effective area files for track-like, muon-induced events.

Using the best-fit VHE energy spectrum from from \cref{tab:fit} and the IceCube IC-86 effective areas for track-like events for the year 2012 from \cite{IceCubeData}, the expected detection rate for muon neutrinos from \theSource{} in the IceCube detector is \SI[scientific-notation=false] {0.58}{\per\year}, or about six per decade, with a median neutrino energy of about \SI{6}{\TeV}. While this is likely not a strong enough signal to be picked up by IceCube, future upgrades to the detector \citep{Aartsen:2014njl} or next-generation neutrino observatories (e.g. \cite{Aiello:2018usb}) might be able to detect neutrino emission from this nearby PeVatron.

\subsection{Leptonic Origin}
If, on the other hand, the observed VHE gamma-ray emission is of leptonic origin, there must be a source nearby that is capable of accelerating electrons to hundreds of TeV. The observed gamma-ray flux levels require at least \SI[scientific-notation=true, exponent-to-prefix=false]{4e45}{\erg} in electrons above \SI[scientific-notation=engineering, exponent-to-prefix=true]{2}{\TeV}, which the supernova shell should be able to provide.

Another possible source of electrons could be the pulsar PSR J2229+6114 and/or its PWN G106.65+2.96. PSR J2229+6114 has a spindown luminosity of \SI[scientific-notation=true, exponent-to-prefix=false]{2.2e37}{\erg\per\second} \citep{Halpern_2001}. As discussed earlier, it is not clear that its characteristic age of 10.4\,\si{\kilo\year} is a good estimate for the actual age of the system. Additionally, the pulsar has shown several timing glitches and its braking index is not well measured \citep{10.1093/mnras/stw3081}. Still, assuming that the pulsar is at least 1\,kyr old and its energy output has not been less than the currently measured value, the pulsar would have released at least \SI[scientific-notation=true, exponent-to-prefix=false]{7e47}{\erg} of kinetic energy over its lifetime; a sufficient energy budget to produce the necessary population of electrons. However, if the pulsar is the source of the relativistic electrons producing the gamma-ray emission, one would expect this emission to be centered around the pulsar.


\section{Conclusions and Outlook}
\label{conclusions}
The HAWC detection of hard-spectrum gamma-ray emission from the \theSnr{} region up to more than 100 TeV is interpreted in the context of both a hadronic and a leptonic emission model. Assuming a hadronic origin of the VHE gamma-ray emission oberved by HAWC and VERITAS, the cutoff energy in the  underlying proton spectrum is constrained to be above 800\,TeV. This would make the source a \emph{Galactic PeVatron}. The supernova shockwave could have released sufficient energy to account for the observed VHE gamma-ray emission.

So far, no hint for a cutoff has been detected in the VHE gamma-ray energy spectrum of \theSource{}. Due to the finite age and size of the SNR, it must cut off somewhere between \SI[scientific-notation=engineering]{120}{\TeV} (the lower limit to the gamma-ray cutoff energy derived from the joint fit) and the PeV range, depending on the magnetic field and the size of the acceleration region. Detecting this cutoff would be important to improve our understanding of this source and how much it could contribute to the Galactic cosmic-ray population at the `knee' of the cosmic-ray spectrum. 

Due to its large celestial latitude, \theSource{} can only be observed at relatively large zenith angles by HAWC and other observatories at similar latitudes. Future upgrades to HAWC's reconstruction algorithms, such as optimizing the energy estimation at large zenith angles or including the additional `outrigger' tanks installed around the main detector array \citep{Marandon:2019sko}, will improve HAWC's sensitivity --- especially at high energies. Future efforts could also benefit from cross-calibration of HAWC and VERITAS, which would improve the uncertainty on the spectral index and hence allow better contraints on the cutoff energy.

Several next-generation gamma-ray observatories are currently under development or construction. \theSource{} will be outside the field of view of both CTA South \citep{Acharya:2017ttl} and SWGO \citep{Abreu:2019ahw}, both optimized for high energies and the study of Galactic cosmic-ray sources. CTA North, located at \ang{28.7622}N, will be able to observe \theSource{}. However, the northern CTA observatory will be optimized for the detection of extragalactic sources and will have worse sensitivity\footnote{Comparing 50 hours with CTA North to 5 years of HAWC, see \url{https://www.cta-observatory.org/science/cta-performance/}} than HAWC above 20 TeV. CTA North could improve its sensitivity (especially at the highest energies) either by spending more time on this source, or by employing dedicated observing strategies such as observations of the rising or setting source at large zenith angles (see \cite{Peresano:2019rxk}).

LHAASO, the Large High-Altitude Air Shower Observatory \citep{DiSciascio:2016rgi}, is located at a latitude of \ang{29.36} and uses the water Cherenkov technique to detect air showers, similar to HAWC. Once complete, LHAASO is expected to have more than an order of magnitude better sensitivity compared to HAWC for a similar run time. LHAASO will be able to make an important contribution to the understanding of the emission spectrum of \theSource{}.

\acknowledgments

We acknowledge the support from: the US National Science Foundation (NSF); the US Department of Energy Office of High-Energy Physics; the Laboratory Directed Research and Development (LDRD) program of Los Alamos National Laboratory; Consejo Nacional de Ciencia y Tecnolog{\'i}a (CONACyT), M{\'e}xico, grants 271051, 232656, 260378, 179588, 254964, 258865, 243290, 132197, A1-S-46288, A1-S-22784, c{\'a}tedras 873, 1563, 341, 323, Red HAWC, M{\'e}xico; DGAPA-UNAM grants AG100317, IN111315, IN111716-3, IN111419, IA102019, IN112218; VIEP-BUAP; PIFI 2012, 2013, PROFOCIE 2014, 2015; the University of Wisconsin Alumni Research Foundation; the Institute of Geophysics, Planetary Physics, and Signatures at Los Alamos National Laboratory; Polish Science Centre grant DEC-2017/27/B/ST9/02272; Coordinaci{\'o}n de la Investigaci{\'o}n Cient{\'i}fica de la Universidad Michoacana; Royal Society - Newton Advanced Fellowship 180385; Generalitat Valenciana, Spain, grant CIDEGENT/2018/034. Thanks to Scott Delay, Luciano D{\'i}az and Eduardo Murrieta for technical support.

\vspace{5mm}
\facilities{HAWC, VERITAS}

\software{astropy \citep{2013A&A...558A..33A},  
          3ML \citep{threeML}, 
          Naima \citep{naima}
          }

\clearpage

\appendix

\section{Instrument and data analysis}
\label{analysis}
\subsection{The HAWC detector}
HAWC is a large-field-of-view (FoV), ground-based air shower detector array located at \ang{18;59;42}\,N, \ang{97;18;27}\,W, in Mexico. It is optimized for gamma-ray astronomy in the TeV regime. The detector, event reconstruction, binning, and background estimation are described in \cite{2HWC,hawcCrab}. HAWC's energy threshold depends on the source's declination and its energy spectrum, and ranges from hundreds of GeV to tens of TeV. HAWC's angular resolution (68\% containment radius) improves with the fraction of the array triggered by an air shower and the source elevation, and varies from $\sim$\ang{1} to $\sim$\ang{0.2} \citep{hawcCrab}. About two thirds of the sky (8.4 sr, from \ang{-26} to \ang{64} declination) are visible to HAWC, with its $>95\%$ duty cycle and an instantaneous FoV covering about \SI{1.8}{\steradian}. HAWC data corresponding to 1347 sidereal days of livetime are used for this analysis.

\subsection{Data analysis}
\label{sec:data_analysis}
The source search follows the method described in \cite{2HWC}: A putative point source with a power-law energy spectrum (spectral index: $-2.5$) is moved across the sky. The grid points correspond to the centroids of HEALPix pixels \citep{Gorski_2005} with NSIDE=1024. For each source position, the flux normalization is fit and a likelihood ratio of the best-fit source+background model $\hat{\mathcal{L}}_{s+b}$, compared to the background-only model $\mathcal{L}_b$, is calculated. A test statistic (TS) is derived from this likelihood ratio: $TS=2\,\log\left( \hat{\mathcal{L}}_{s+b} / \mathcal{L}_b\right)$. The local maxima of the \emph{significance map} obtained with this procedure are candidate gamma-ray sources (see \cref{fig:sigmap}).  

Following the initial source search, the 3ML\footnote{\url{www.github.com/threeML/}} (multi-mission maximum likelihood) framework \citep{threeML} with the HAL\footnote{\url{www.github.com/threeML/hawc_hal}} (HAWC accelerated likelihood) plugin is used to perform likelihood fits to determine the morphology and energy spectrum of the source. The likelihood calculation within the HAL plugin proceeds similarly to previous HAWC publications \citep{hawcCrab,liff}.


For joint fits with VERITAS data, spectral points from \cite{2009ApJ...703L...6A} are added to the likelihood calculation via a $\chi^2$-like likelihood. VERITAS quotes a Gaussian half-width of \ang{0.27\pm 0.05} by \ang{0.18\pm 0.03} for the spatial extent of the source. Accounting for their angular resolution of \ang{0.1}, about 40\% of their observed emission is expected to fall outside of their source region (a circular region with a radius of \ang{0.32}). For the joint spectral fits, the VERITAS differential flux measurements are scaled up by a factor of 1.67 to account for this effect.

In this study, the differential photon fluxes $dN/\left(dE\,dA\,dt\right)$ are modeled as power laws (PL):
\begin{equation}
\frac{dN}{dE\,dA\,dt} = K \, \left(\frac{E}{E_0}\right)^{\gamma}. \label{eq:PL}
\end{equation}
Power-law spectra with an exponential cutoff (CPL) are also tested:
\begin{equation}
\frac{dN}{dE\,dA\,dt} = K \, \left(\frac{E}{E_0}\right)^{\gamma} \, \exp\left( -\frac{E}{E_C}\right). \label{eq:CPL}
\end{equation}

Here, $E$ is the (true) gamma-ray energy, $K$ is the flux normalization, $\gamma$ the spectral index, $E_0$ the pivot energy, and $E_C$ is the cutoff energy. The gamma-ray emission is assumed to be isotropic and constant in time over the last two decades.

As described in \cite{hawcCrab}, HAWC uses the fraction of photon detectors measuring a signal ($f_{hit}$) as a proxy for the gamma-ray energy. $f_{hit}$ is also correlated with the zenith angle of a given shower and its position with respect to the array, making it non-trivial to recover the energy of a given gamma-ray shower. Per-event energy estimators, introduced in \cite{EEst}, are not used here due to the high declination of the source. To calculate the energy range over which the spectral fits to HAWC data are valid, the hard-cutoff method presented in \cite{hawcCrab} is used.

Lower limits on the exponential cutoff energy $E_C$ are determined from a likelihood ratio test. The likelihood of a model with an exponential cutoff in the energy spectrum, $\hat{\mathcal{L}}_{CPL}$, is compared to the best-fit model with no cutoff, $\hat{\mathcal{L}}_{PL}$. A test statistic is derived from this: $TS=2\, \left( \log \hat{\mathcal{L}}_{CPL} - \log \hat{\mathcal{L}}_{PL} \right)$. We then estimate the 90\% confidence interval on $E_C$ by scanning over a range of cutoff energies, re-optimizing the spectral parameters of the cutoff model for each point in the scan, and identifying the value $E_C$ where $TS=-1.64$.

\subsection{Modeling of the underlying particle population}

\subsubsection{Hadronic modeling}
The observed VHE emission can be interpreted in the context of a hadronic emission model: Relativistic protons interact with ambient hydrogen nuclei, producing a cascade of particles including neutral pions, which decay into gamma rays. The Naima framework \citep{naima} is used to predict the resulting gamma-ray emission spectrum from a given proton population. The parameters of the underlying proton spectrum are fit to the data. 

Two spectral models are tested for the proton energy spectrum $dN_p/dE_p$: a simple power law (\cref{eq:PL}) and a power law with a cutoff (\cref{eq:CPL}). Instead of the normalization $K$ at a given pivot energy, the total proton energy $W_p$ is used to describe the normalization of the proton spectrum:

\begin{equation}
W_p = \int \limits_{E_1}^{E_2} E_p\, \frac{dN_p}{dE_p} dE_p.
\end{equation}

$E_1$ and $E_2$ are the minimum and maximum energies of the proton population used to predict the gamma-ray spectrum. 

Care should be taken in selecting the energy range; choosing a too-small energy range, meaning $E_1$ too high (or $E_2$ too low), will cause Naima to underestimate the gamma-ray flux at low (high) energies. On the other hand, choosing a too-large range, meaning $E_1$ too low (or $E_2$ too high), can increase the overall uncertainty on $W_p$ and the correlation between $W_p$ and the spectral index, as we might be extrapolating the proton spectrum to energies where it is unconstrained by our measurements. We chose $E_1 = \SI[scientific-notation=engineering]{2}{\TeV}$ as a compromise where the flux over the gamma-ray energy range used here (0.9\,TeV to 180\,TeV) is underestimated by at most 5\% and $E_2 = \SI[scientific-notation=engineering]{1e15}{\keV}$.

For a given spectral shape of the proton spectrum, the predicted gamma-ray emission is proportional to $n\, W_p/d^2$, where $n$ is the gas density in the emission region and $d$ is the distance between the observer and the source. Accordingly, two of these three parameters have to be fixed during the fit procedure. In this study, $d$ is fixed to the measured value of \SI[exponent-to-prefix=false, scientific-notation=false]{800}{\parsec} \citep{Kothes_2001}, and the best-fit normalization is reported in terms of $n\, W_p$.


\subsubsection{Leptonic modeling}

Relativistic electrons and positrons (from hereon, electrons) emit electromagnetic radiation via three main processes (see e.g. \cite{RevModPhys.42.237} for more details): synchrotron emission due to deflection by ambient magnetic fields, bremsstrahlung due to scattering with ambient ions/nuclei, and inverse Compton-upscattering of lower-energy photons (IC emission). In most cases, TeV gamma-ray emission from electrons is dominated by the IC process. Using Naima, we model the observed VHE emission from \theSource{} as IC emission from three seed photon fields: the cosmic microwave background, a galactic near-infrared photon field, and a galactic far-infrared photon field. For all three fields, the default values set by the Naima package are used. As in the hadronic model, the energy spectrum of the electrons is modeled as a power-law spectrum (\cref{eq:PL}) between 2\,TeV and 1\,EeV, with the index $\gamma_e$ and the total electron energy $W_e$ being the only two free parameters of the fit. A power-law spectrum with an exponential cutoff (\cref{eq:CPL}) was tested as well. 

\subsection{Systematic uncertainties}
Two classes of systematic uncertainty are considered here: the modeling of the HAWC instrument response and the uncertainty of the VERITAS measurements. Uncertainties related to the HAWC detector model are investigated as in \cite{EEst}. To incorporate the systematic uncertainties on the VERITAS data, the joint fit is repeated four more times, with the VERITAS data points adjusted according to the quoted systematic uncertainty: all fluxes scaled up/down by the constant factor $\left(1\pm \Delta K/K\right)$, and scaled individually by an energy-dependent factor $\left(E/E_0\right)^{\pm \Delta\gamma}$. Here, $E_0$ designates the pivot energy, $\Delta K$ is the systematic uncertainty on the flux normalization $K$, and $\Delta \gamma$ is systematic uncertatinty on the spectral index. The resulting shifts in the fit parameters were added in quadrature, as in the ``shift method'' from \cite{doi:10.1146/annurev.nucl.57.090506.123052}.



\section{Molecular Gas in the Region}
\label{molecular_gas}

Here, we briefly outline how the density of molecular gas in the \theSnr{} region is determined. 

We used data from the DAME CO survey \citep{Dame_2001}, available online\footnote{See \url{https://www.cfa.harvard.edu/rtdc/CO/CompositeSurveys/}. The dataset used here is the interpolated whole Galaxy ``Local'' Cube, \url{https://www.cfa.harvard.edu/rtdc/CO/download/COGAL_local_interp.fits.gz}}.

The available dataset contains the brightness temperature $T$ of the $^{12}{\rm C}^{16}{\rm O} ~ J=1\rightarrow 0$ line, with a velocity resolution of \SI[scientific-notation=engineering]{1.3}{\kilo\metre\per\second} and a grid spacing of \ang{0.125}, potentially interpolated from sparser measurements.

Scanning a square of width \ang{0.6} around the nominal position of \theSource{}, we find a peak in the brightness temperature at \SI[scientific-notation=engineering]{-5.2}{\kilo\metre\per\second}, presumably corresponding to the SNR. For the gas maps of the region, we therefore consider the three velocity bins centered at \SI[scientific-notation=engineering]{-6.5}{\kilo\metre\per\second}, \SI[scientific-notation=engineering]{-5.2}{\kilo\metre\per\second}, and \SI[scientific-notation=engineering]{-3.9}{\kilo\metre\per\second}.

The column density of hydrogen molecules can be determined from the measured temperature brightness as 

\begin{equation}
N_{H_2} = X_{CO} \, \int\limits_{v_{min}}^{v_{max}} T(v) dv.
\end{equation} 

The CO-to-H$_2$ conversion factor is given by $X_{CO}\approx 2\cdot 10^{20}\,\si{\per\centi\metre\squared \per\kelvin \per\kilo\metre \second}$ according to \cite{doi:10.1146/annurev-astro-082812-140944}. Here, the integration is replaced by the summation over the previously indicated velocity bins. 

\cref{fig:sigmap} shows the column density of hydrogen molecules determined in this way. There is a cloud of molecular gas centered around RA=\ang{336.8}, Dec=\ang{60.85}, consistent with the VHE emission region. Integrating the molecular hydrogen in a circle of radius \ang{0.3} around this point yields a gas content of roughly 400 $M_\odot$. Assuming a spherical region of radius $\SI[scientific-notation=engineering]{800}{\parsec} \cdot \tan\left(\ang{0.3}\right) \approx \SI[scientific-notation=engineering]{4}{\parsec}$, this corresponds to an average density of hydrogen \emph{atoms} of about \SI[scientific-notation=false, exponent-to-prefix=false]{50}{\per\cubic\centi\metre}, which we adopt in the calculation of the proton energy needed to produce the observed gamma-ray emission. We assume here that the molecular gas content dominates the emission region over atomic or ionized hydrogen.

\clearpage
\bibliographystyle{aasjournal}
\bibliography{paper}



\end{document}